# 3D optoelectronics and co-packaged optics: when solving the wrong problems stalls deployment

Yasha Yi[1,2*] and Danny Wilkerson[3]

[1]*Intelligent Optoelectronics Laboratory and Connected Systems Institute, University of Wisconsin, Milwaukee, WI 53211*

[2]*Microsystems Technology Laboratory, Massachusetts Institute of Technology, Cambridge, MA 02139*

[3] *Invictus Innovation EV Technology, Detroit, MI 48098*

*[yiy@uwm.edu](yiy@uwm.edu) or [yys@alum.mit.edu](yys@alum.mit.edu)*

## Abstract

The rapid growth of AI and accelerator-driven workloads is forcing a fundamental rethinking of optical interconnect architectures in datacenters. Co-packaged optics and three-dimensional photonic integration have emerged as promising solutions to overcome the energy and bandwidth limitations of electrical I/O. Yet, as optics move closer to compute, packaging, thermal management, and system-level robustness increasingly dominate performance and scalability. Here, we argue that co-packaged optics should not be viewed as a component-level optimization, but as an architectural commitment that reshapes the boundaries between photonics, electronics, and system design. We examine how heterogeneous integration strategies, chiplet-based optics, and emerging packaging platforms redefine scaling laws for AI systems, often introducing trade-offs that are underappreciated in device-centric analyses. Looking forward, we discuss why standardization, serviceability, and thermal-aware co-design will be decisive in determining whether co-packaged optics can transition from early deployment to widespread adoption in AI-scale datacenters.

## 1. The AI datacenter has outgrown optical I/O assumptions

The rapid rise of artificial intelligence (AI) workloads is reshaping datacenter design at a pace that is outstripping long-standing assumptions about optical input–output (I/O). For decades, optical interconnects were treated as peripheral enablers—high-bandwidth, energy-efficient links that could be upgraded independently of compute. That abstraction is now breaking down. As AI accelerators scale in both number and power density, optical

I/O is no longer a separable subsystem but a structural element that constrains architecture, packaging, and system reliability.

Traditional optical I/O paradigms emerged in an era dominated by relatively modular server architectures and predictable traffic patterns. Bandwidth scaling was achieved by increasing lane counts and data rates, while optics remained physically and conceptually distant from compute silicon. Energy per bit and reach were the dominant metrics, and pluggable transceivers offered a convenient balance between performance, flexibility, and serviceability. These assumptions held as long as electrical I/O could bridge the gap between switches, accelerators, and optical modules without dominating power or signal integrity budgets [1-7].

AI datacenters invalidate these premises. Modern training and inference workloads are defined by dense, low-latency, all-to-all communication across thousands of accelerators. The resulting traffic is highly localized, persistent, and intolerant of latency variability. Electrical I/O has become a primary bottleneck, forcing optics to migrate ever closer to the compute die. Co-packaged optics (CPO) and three-dimensional (3D) photonic integration represent natural responses to this pressure [8-13], reducing electrical trace lengths and lowering energy per bit. Yet these solutions also collapse the historical separation between optics, electronics, and system design.

This collapse exposes a deeper mismatch between AI system requirements and inherited optical I/O assumptions. Energy efficiency, while still important, is no longer the sole—or even dominant—optimization target. In AI clusters, incremental gains in pJ per bit can be negated by thermal instability, yield loss, or operational fragility at the package or rack level. As optical engines are placed adjacent to, or stacked with, multi-hundred-watt accelerator dies, thermal gradients, mechanical stress, and wavelength drift become first-order system concerns. Optical performance can no longer be specified independently of cooling strategy, power delivery, or packaging topology.

Early demonstrations of CPO have understandably emphasized headline metrics: reduced energy per bit, increased aggregate bandwidth, and compact form factors. These results validate the physical advantages of short-reach optical I/O, but they also mask emerging system-level constraints. In tightly integrated AI nodes, the failure or degradation of a single optical interface can compromise an entire package, rather than a replaceable module. Yield losses compound across heterogeneous dies, and maintenance models designed around hot-swappable optics no longer apply [14-31]. The historical assumption that optics are serviceable endpoints at the edge of a system is increasingly incompatible with AI-scale integration [32].

Moreover, AI workloads amplify the consequences of architectural rigidity. Accelerator lifecycles are shortening, interconnect topologies are evolving rapidly, and workload characteristics vary across training, inference, and emerging hybrid models. Optical I/O architectures optimized for a specific generation of switches or accelerators risk premature obsolescence if they lack modularity or adaptability. In this context, the question is not simply how to integrate optics closer to compute, but how to do so without hard-coding today's assumptions into tomorrow's infrastructure.

These pressures reveal a fundamental shift: optical I/O has transitioned from a bandwidth provisioning problem to a system co-design problem. Decisions about photonic platform, packaging strategy, and integration depth now directly influence datacenter-level metrics such as uptime, scalability, and total cost of ownership. For AI systems operating at megawatt scale, robustness per watt may be more consequential than energy per bit. Optical links that are marginally less efficient but thermally stable, manufacturable at scale, and compatible with chiplet-based replacement strategies may ultimately deliver higher system-level performance.

The AI datacenter, therefore, is not merely stressing existing optical technologies—it is redefining the criteria by which optical I/O solutions should be evaluated. Architectures that prioritize component-level optimization without accounting for heterogeneous integration, thermal coupling, and lifecycle management risk solving system's problems. As optics continue their migration into the package, and increasingly into the third dimension, the field must confront an uncomfortable reality: the success of co-packaged and 3D photonic systems will be determined less by optical elegance than by architectural restraint.

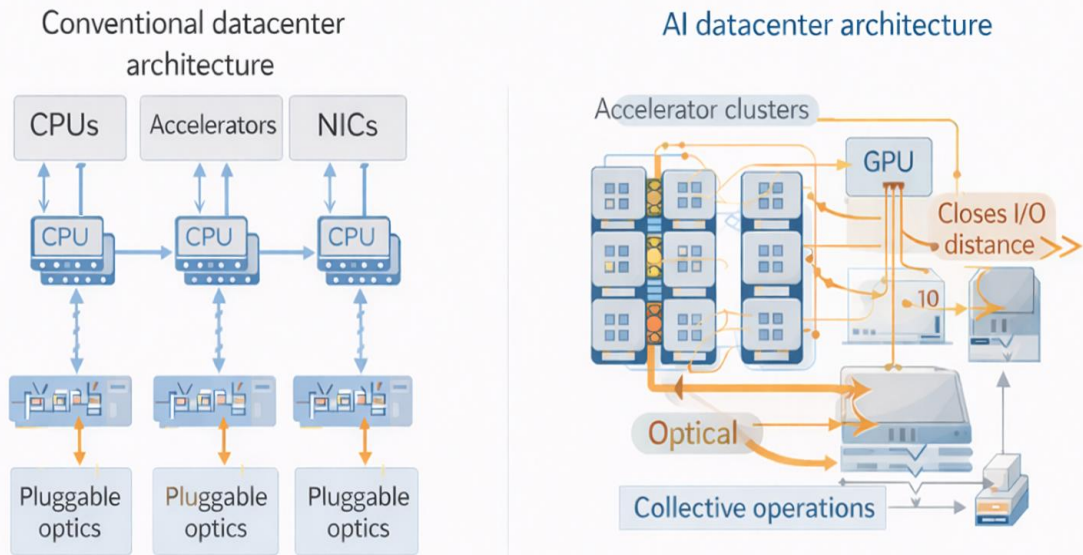

Figure 1 | AI datacenter workloads invalidate legacy optical I/O assumptions.

Conventional datacenter architectures (left) rely on electrical interconnects for short-reach communication and deploy optics primarily at the board or system boundary. In contrast, AI datacenters (right) are dominated by accelerator-centric, high-bandwidth, all-to-all communication patterns that compress latency budgets and electrical reach, progressively pulling optical I/O closer to the compute package.

In this new regime, optical I/O can no longer be designed in isolation. It must be conceived as an integral, system-aware element of AI computing platforms—one whose limitations shape, and are shaped by, the broader architecture of the datacenter itself.

Much of the current discourse around co-packaged optics understandably emphasizes feasibility, performance milestones, and near-term deployment[33-34]. Far less attention has been paid to how deeply embedding optics within compute packages redefines system boundaries, lifecycle management, and innovation pathways at datacenter scale. Impressive demonstrations validate feasibility. The harder question is whether these architectures remain operationally resilient at hyperscale.

**2. Co-packaged optics is not a component — it is a system decision**

As optical interconnects move from the periphery of the datacenter into the package, co-packaged optics (CPO) can no longer be treated as a drop-in component upgrade. Instead, CPO represents a system-level decision that reshapes architectural boundaries across compute, packaging, cooling, and operations. This distinction is subtle but consequential:

components can be optimized locally and replaced independently, whereas systems impose constraints that propagate across multiple design layers and lifecycles.

Historically, optical transceivers occupied a well-defined role. They interfaced standardized electrical I/O on one side and optical fibers on the other, enabling system architects to abstract optics away from compute and switching logic. Performance tuning focused on link-level metrics—bit rate, reach, and energy per bit—while failure modes were largely localized and serviceable. Co-packaged optics disrupts this abstraction. By placing optical engines in close physical and thermal proximity to high-power ASICs, CPO collapses once-independent design domains into a tightly coupled system.

This coupling transforms the nature of optimization. In a CPO architecture, optical performance is inseparable from electrical signaling, thermal management, and mechanical integrity. Electrical trace length is reduced, but thermal gradients increase. Optical alignment tolerances improve at the die level, yet assembly yield becomes sensitive to multi-die interactions. Decisions that might appear favorable from a photonic perspective—such as higher modulation density or tighter wavelength spacing—can introduce instability when embedded in a thermally dynamic environment dominated by AI accelerators operating at hundreds of watts.

The implications are particularly acute in three-dimensional (3D) integration schemes. Vertical stacking promises the shortest interconnects and highest bandwidth density, but it also amplifies non-linear interactions between layers. Heat generated by logic dies directly perturbs photonic elements below or above them, affecting laser efficiency, resonator stability, and link margins. Unlike planar systems, where thermal paths can be engineered laterally and mitigated with relative flexibility, 3D stacks constrain heat flow in ways that are difficult to retrofit. In such configurations, optical I/O becomes a thermally sensitive subsystem embedded within the most power-dense region of the package.

These challenges reveal a broader architectural truth: the benefits of CPO cannot be evaluated in isolation from their systemic costs. Energy per bit reductions achieved by eliminating long electrical traces may be offset by the energy overhead required to stabilize temperatures, tune wavelengths, or manage thermal crosstalk. Similarly, gains in bandwidth density may come at the expense of yield, reliability, or serviceability. The question is no longer whether CPO is technically feasible, but whether specific implementations are architecturally sustainable at scale.

The distinction between silicon photonics–based and VCSEL-based CPO platforms illustrates this point. These approaches are often compared in terms of reach, modulation efficiency, or wavelength-division multiplexing capability. However, in a system context, they represent fundamentally different architectural commitments. Silicon photonics enables dense integration and longer reach but demands stringent alignment, thermal control, and often external or heterogeneously integrated light sources. VCSEL-based solutions, by contrast, offer simpler coupling and exceptional short-reach efficiency, but rely on parallelism and face limitations in scaling wavelength density. Neither approach is inherently superior; their viability depends on how their constraints interact with packaging, cooling, and operational models at the system level.

Crucially, CPO also alters failure semantics. In traditional architectures, optical transceivers are among the most frequently replaced components in a datacenter. Their pluggability enables maintenance without disturbing the surrounding system. Co-packaged optics erodes this separation. When optical engines are embedded within the same package as a switch or accelerator, failures become more consequential and less localized. A degraded optical interface may necessitate replacing an entire package or board, raising costs and complicating operational workflows. While emerging chiplet and connectorized approaches aim to mitigate this risk, they introduce additional interfaces whose reliability must be proven at scale.

These considerations underscore why CPO should be viewed as a system decision rather than a component upgrade. Once optics are co-packaged, choices about integration depth, packaging topology, and optical platform constrain not only performance but also manufacturability, testability, and lifecycle management. Architectural flexibility—the ability to adapt interconnect strategies as workloads and technologies evolve—becomes as important as peak efficiency.

For AI datacenters, where hardware generations turn over rapidly and workloads evolve unpredictably, architectural rigidity poses a significant risk. Systems optimized narrowly around a specific CPO implementation may struggle to accommodate changes in accelerator design, interconnect topology, or cooling infrastructure. Conversely, architectures that acknowledge CPO as a system-level commitment can be designed with modularity, redundancy, and adaptability in mind, even if this entails accepting modest compromises in component-level metrics.

Ultimately, the central challenge is not integrating optics into packages, but integrating optical thinking into system architecture. Co-packaged optics forces designers to confront trade-offs that were previously hidden behind modular interfaces. It demands a shift from optimizing isolated components to balancing interacting subsystems under real-world constraints. As AI computing platforms continue to scale, the success of CPO will depend less on how closely optics can be placed to silicon, and more on how thoughtfully they are embedded into the broader system fabric.

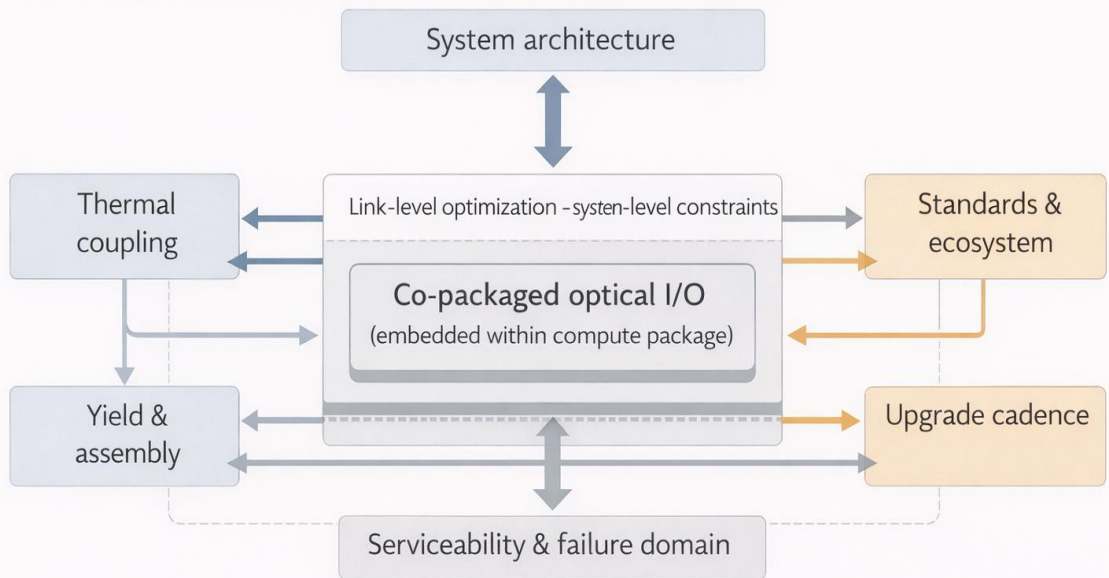

**Figure 2** | Co-packaged optics introduces system-level constraints

Embedding optical I/O within the compute package couples photonic integration to thermal coupling, yield and assembly, upgrade cadence, system architecture, and ecosystem standardization, resulting in system-level architectural constraints beyond the traditional link-level concerns.

In this sense, co-packaged optics is not merely a technological evolution—it is an architectural inflection point. Whether it enables sustainable scaling or introduces new fragilities will depend on how clearly the field recognizes that optics, once co-packaged, cease to be a component and become a defining element of the system itself.

**Table 1** | System-level consequences of integrating optical I/O within the compute package

| Dimension | Pluggable optics | Co-packaged optics |
|---|---|---|
| Replaceability | Hot-swappable modules | Fixed to the compute package |
| Failure domain | Link-level or module-level | Package-or system-level |
| Yield ownership | Optical module vendor | System integrator / platform owner |
| Upgrade cadence | Independent of compute | Coupled to compute and packaging cycle |
| Thermal coupling | Weak (thermally isolated) | Strong (shared thermal environment) |
| Standardization pressure | Moderate | **Critical** |
| Architectural flexibility | High | Reduced, tightly constrained |

Moving optical interconnects from pluggable modules to co-packaged implementations fundamentally alters failure domains, yield ownership, upgrade cycles, and standardization requirements, highlighting the architectural implications of optical proximity beyond link-level performance.

In practice, this transition is unlikely to be binary: intermediate architectures that place optical engines near—but not within—the compute package can preserve short electrical reach while mitigating thermal coupling and serviceability constraints, underscoring the continuum of system-level trade-offs between pluggable and fully co-packaged approaches. Demonstrations of impressive bandwidth and efficiency validate the physical promise of co-packaging. The more difficult question is whether these architectures remain serviceable, upgradeable, and economically predictable across tens of thousands of nodes.

### 3. Packaging and heterogeneous integration now dominate scaling laws

As co-packaged optics migrates from prototype demonstrations to AI-scale deployment, it becomes increasingly clear that classical scaling metrics—bandwidth per lane, energy per bit, or modulation efficiency—are no longer sufficient predictors of system viability. Instead, the dominant constraints are set by packaging and heterogeneous integration. These layers now define the effective scaling laws of optical I/O, determining not only how much bandwidth can be delivered, but how reliably, manufacturably, and sustainably it can be deployed at scale.

Historically, advances in optical interconnects were governed by device-level improvements. Faster modulators, lower-loss waveguides, and more efficient light sources translated relatively directly into system gains. Packaging served primarily as an enabling layer, responsible for electrical routing and mechanical support. In AI datacenters, this hierarchy has inverted. As optical engines are embedded within or adjacent to compute packages, packaging is no longer passive infrastructure; it actively constrains thermal behavior, yield, integration density, and serviceability. These constraints scale nonlinearly with system size, often overwhelming incremental gains achieved at the device level.

One manifestation of this shift is the growing dominance of yield compounding in heterogeneous systems. Co-packaged optical modules typically integrate multiple dies—logic, drivers, photonics, and sometimes light sources—each fabricated in different process nodes and foundries. While individual die yields may be high, the effective yield of the assembled system is their product. As integration density increases, even small variations in alignment tolerance, bonding quality, or thermal stress can lead to disproportionate losses. At AI-datacenter volumes, such yield penalties translate directly into cost, availability, and deployment risk, effectively capping the practical scale of integration.

Thermal scaling further illustrates how packaging governs system limits. In advanced AI accelerators, power densities are already pushing the limits of conventional cooling. Introducing optical engines into this environment does not simply add heat; it redistributes thermal gradients in ways that affect optical stability. Resonant photonic devices are sensitive to temperature fluctuations on the order of a few degrees, while laser efficiency and wavelength drift are tightly coupled to local thermal conditions. Packaging architectures that concentrate heat vertically, as in aggressive three-dimensional stacks, can therefore impose hard limits on optical density long before device performance is exhausted. In this regime, the maximum usable bandwidth is set not by photonic capability, but by the ability of the package to maintain thermal equilibrium under dynamic workloads.

These effects expose a critical distinction between two-dimensional, 2.5D, and fully three-dimensional integration. While vertical stacking offers compelling gains in interconnect length and footprint, it also compresses thermal paths and mechanical tolerances. Lateral integration strategies—such as interposers, embedded bridges, or fan-out architectures—trade some electrical and optical compactness for improved thermal spreading and manufacturability. Recent analyses of deeply integrated compute systems illustrate how rapidly thermal and packaging constraints can escalate with integration depth, reinforcing the need for architecture-aware trade-offs rather than maximal proximity. From a scaling-law perspective, these approaches often exhibit more graceful degradation: bandwidth density increases more slowly, but yield and reliability scale more predictably. For AI datacenters prioritizing uptime and lifecycle cost, such trade-offs may prove advantageous.

Heterogeneous integration also reshapes the economics of scaling. Advanced packaging techniques introduce costs that do not scale linearly with performance. Fine-pitch interposers, precision bonding, and complex assembly flows raise the marginal cost of each additional terabit of bandwidth. At the same time, the operational cost of failures increases as optics become less accessible and more tightly coupled to compute. These economic scaling laws favor architectures that balance integration depth against modularity, even if this means accepting less aggressive physical proximity between optics and logic.

The dominance of packaging constraints also reframes the role of standardization. In earlier generations of optical I/O, standards focused on electrical interfaces and optical form factors, enabling interoperability at the module level. In co-packaged systems, the critical interfaces shift inward, toward die-to-die connections, optical I/O footprints, and thermal-mechanical boundaries. Without standardized approaches to these interfaces, each CPO implementation becomes a bespoke solution, limiting ecosystem participation and slowing adoption. From a scaling perspective, lack of standardization acts as friction, increasing the effective cost and risk of deployment as system size grows.

Importantly, these constraints are not merely engineering challenges to be solved incrementally. They reflect fundamental interactions between physics, manufacturing, and system architecture. Attempts to push integration density without addressing packaging-dominated scaling laws risk encountering abrupt failure modes rather than smooth performance saturation. In AI datacenters, where deployments involve tens of thousands of nodes, such failure modes are unacceptable. Scaling must be predictable, not just theoretically achievable.

Recognizing packaging and heterogeneous integration as primary scaling determinants leads to a different optimization philosophy. Instead of maximizing local metrics—such as bandwidth per square millimeter—architects must consider global metrics, including yield-adjusted bandwidth, thermally stable throughput, and serviceable performance per rack. These metrics inherently favor designs that are robust to variation, tolerant of thermal dynamics, and compatible with iterative deployment and replacement. In this framework, the most successful optical architectures may not be the most densely integrated, but the ones that scale most gracefully under real-world constraints.

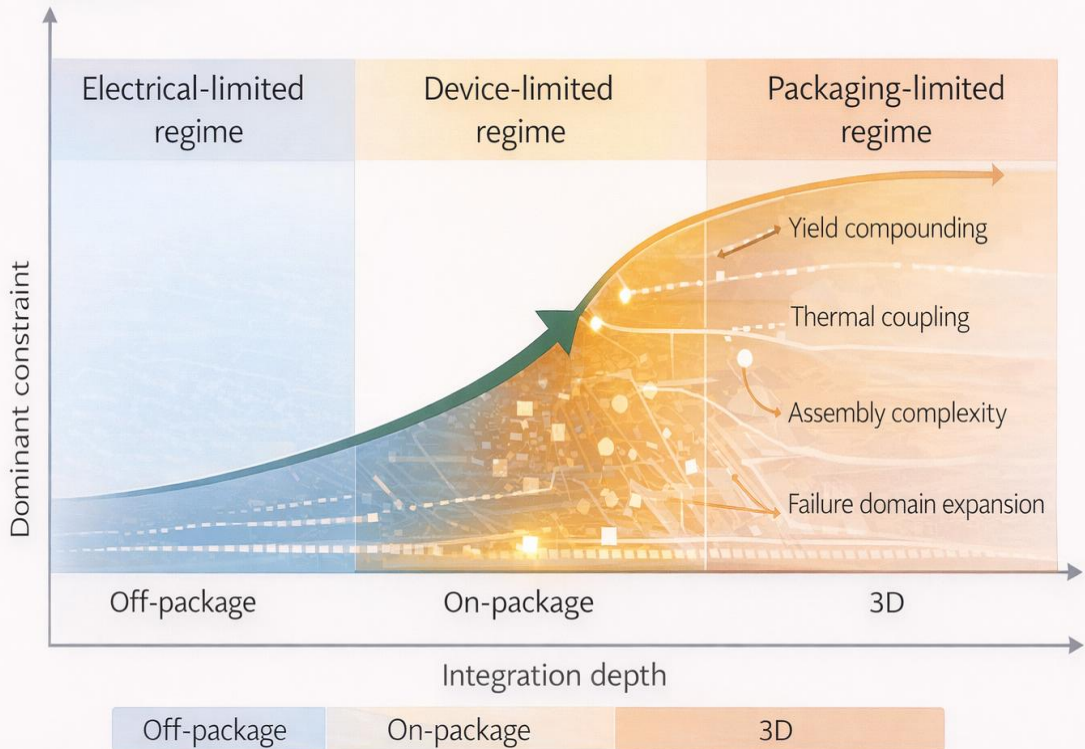

Figure 3 | Packaging and heterogeneous integration redefine system-level scaling limits for optical I/O.

As optical I/O moves from board-level implementations toward in-package and three-dimensional integration, dominant constraints shift from electrical reach and device efficiency to yield compounding, thermal coupling, and assembly complexity, making packaging—rather than photonic device performance—the primary limiter of system-

In this sense, packaging has become the new "process node" for optical I/O. Just as transistor scaling once dictated the trajectory of electronic systems, packaging and heterogeneous integration now define the feasible envelope for co-packaged optics in AI datacenters. Progress will depend less on isolated breakthroughs in photonic devices than on coordinated advances in assembly, thermal management, and interface standardization. Until these layers mature, the scaling of optical I/O will remain bounded not by what photonics can do, but by what packaging can reliably sustain.

**4. Chiplet optics and standardization will determine adoption, not performance**

Taken together, these considerations suggest three architectural guardrails for co-packaged optics at AI scale: (i) integration depth should scale with thermal stability rather than bandwidth density alone; (ii) yield-adjusted bandwidth, not peak bandwidth, should guide packaging decisions; and (iii) modularity must be preserved wherever accelerator and optical innovation cycles diverge. As co-packaged optics transitions from laboratory demonstrations to deployment in AI datacenters, the factors that govern adoption shift decisively away from peak performance metrics. Energy per bit, bandwidth density, and modulation efficiency remain necessary conditions, but they are no longer sufficient. At scale, adoption hinges on whether optical interconnects can be modularized, interoperable, and integrated into an ecosystem that tolerates rapid iteration. In this context, chiplet-based optics and standardization emerge not as conveniences, but as prerequisites.

The appeal of chiplet optics lies in their ability to decouple innovation cycles. AI accelerators, switch ASICs, and optical engines evolve on different timelines, are fabricated in different process nodes, and face distinct yield and cost pressures. Monolithic integration promises elegance and compactness, but it ties these domains together in ways that magnify risk. A change in optical requirements can necessitate a redesign of the entire package; a yield issue in one component can stall deployment across the system. Chiplet-based approaches mitigate these risks by allowing optical functions to be developed, tested, and upgraded independently before assembly.

This decoupling is particularly valuable in AI datacenters, where architectural assumptions change rapidly. Training models grow in size, interconnect topologies evolve, and accelerator designs are refreshed frequently. Optical I/O that is tightly bound to a specific generation of compute risks becoming a bottleneck rather than an enabler. Chiplet optics, by contrast, support incremental evolution. Optical engines can be optimized for specific reach, bandwidth, or power targets and paired with different compute dies as system requirements shift. Even when co-packaged, the logical modularity introduced by chiplets preserves a degree of architectural flexibility that monolithic designs struggle to offer.

However, chipletization alone does not guarantee scalability. Without standardization, chiplet-based optics risk devolving into bespoke integrations that replicate the fragmentation of earlier optical ecosystems. In AI datacenters, where deployment volumes are high and margins are thin, proprietary interfaces act as barriers to adoption. They limit supplier diversity, complicate validation, and slow the pace at which new solutions can be introduced. Standardization, therefore, is not merely about convenience—it is about reducing systemic friction.

Crucially, the locus of standardization must move inward. Traditional optical standards focus on pluggable form factors, electrical interfaces, and link specifications at the system edge. Co-packaged and chiplet-based optics require a different emphasis. The critical interfaces now include die-to-die electrical connections, optical I/O footprints, fiber attach geometries, and thermal-mechanical boundaries within the package. These interfaces determine whether optical chiplets from different suppliers can be integrated predictably and whether systems can be assembled and tested at scale.

Packaging design kits and assembly standards play a central role in this shift. While photonic process design kits have accelerated device-level innovation, they do little to address the challenges of heterogeneous integration. For AI datacenters, where reliability and yield dominate cost models, standardized packaging abstractions are arguably more important than standardized photonic devices. They enable system designers to reason about thermal paths, alignment tolerances, and signal integrity without resorting to bespoke co-design for each implementation. In effect, they provide the "contract" that allows chiplet ecosystems to function.

The emphasis on standardization also reframes how performance should be evaluated. In isolation, a highly optimized optical engine may demonstrate exceptional metrics. In a standardized, chiplet-based system, performance must be assessed in terms of compatibility, repeatability, and lifecycle cost. An optical solution that delivers slightly lower bandwidth density but integrates seamlessly into a standardized package may outperform a superior device that requires custom assembly or tuning. For AI datacenters operating at scale, predictable behavior often outweighs theoretical optimality.

Serviceability further reinforces this dynamic. Traditional datacenter operations rely on modular replacement and staged upgrades. Co-packaged optics challenges these practices by embedding optical functionality deep within packages. Chiplet architectures offer a partial remedy by enabling localized replacement or reconfiguration, but only if interfaces are standardized and accessible. Without such standards, even chiplet-based optics risk becoming effectively monolithic from an operational perspective, undermining one of their key advantages.

The tension between performance-driven design and ecosystem-driven adoption is not unique to optics, but it is particularly acute in AI systems. Accelerator performance gains are often measured in multiples, while optical interconnect improvements are incremental and constrained by physics. In this environment, the relative importance of ecosystem efficiency grows. Time to deployment, supplier diversity, and ease of integration can outweigh marginal gains in link efficiency. Optical technologies that align with these priorities are more likely to see widespread adoption, even if they sacrifice some performance headroom.

Ultimately, the success of co-packaged optics in AI datacenters will depend less on how aggressively optics can be integrated, and more on how effectively they can be standardized and modularized. Chiplet-based architectures provide a path toward balancing integration with flexibility, but only if they are supported by robust, widely adopted standards that address packaging and assembly as first-class concerns. In the absence of such standards, performance advances risk remaining confined to niche deployments.

In this sense, the decisive competition is not between photonic platforms, but between ecosystems. Optical I/O solutions that enable interoperability, reduce integration risk, and support rapid iteration will scale. Those that optimize performance at the expense of modularity and standardization may achieve impressive demonstrations, but struggle to transition into the operational reality of AI datacenters.

## 5. What success looks like in 5–10 years

If co-packaged optics and three-dimensional photonic integration succeed over the next decade, their impact will be measured less by record-breaking demonstrations than by quiet architectural normalization. Success will not look like optics disappearing into ever-denser packages at all costs; rather, it will resemble a mature equilibrium in which optical I/O is treated as a first-class system resource—integrated where it adds value, decoupled where it preserves flexibility, and engineered with lifecycle considerations on par with performance.

In practical terms, successful AI datacenters will deploy heterogeneous optical architectures rather than a single, monolithic solution. Co-packaged optics will be used selectively at the most bandwidth- and latency-critical interfaces, such as accelerator-to-switch or intra-rack fabrics, while less demanding links retain more modular optical form factors. This hybridization reflects a recognition that integration depth is not an absolute good, but a context-dependent trade-off. The most effective systems will combine short-reach, high-density optical interfaces near compute with packaging strategies that avoid excessive thermal coupling and preserve serviceability.

Thermal-aware design will be a defining feature of mature deployments. Instead of treating thermal mitigation as an auxiliary challenge, future systems will incorporate thermal behavior into architectural planning from the outset. Optical engines will be placed, spaced, and operated based on workload-driven thermal profiles, rather than fixed assumptions about steady-state operation. Success in this regime will not necessarily mean achieving the lowest possible energy per bit, but delivering thermally stable throughput across realistic, time-varying AI workloads. Optical I/O that performs consistently under these conditions will be valued more highly than designs that excel only under idealized benchmarks.

Modularity will also be a hallmark of success. In five to ten years, optical chiplets are likely to be treated analogously to compute or memory chiplets today: standardized building blocks that can be sourced, integrated, and upgraded independently. This does not imply a return to fully pluggable optics, but rather the emergence of well-defined internal interfaces that allow optical functionality to evolve without forcing wholesale system redesigns. Such modularity will be essential for coping with the rapid evolution of AI accelerators and interconnect topologies, which show no signs of stabilizing.

Equally important will be the maturation of packaging and assembly ecosystems. Successful optical systems will rely on packaging platforms that are not only high performance, but also repeatable, testable, and scalable in volume. Yield-aware design practices will be embedded into architectural decisions, reducing sensitivity to small variations in alignment, bonding, or thermal stress. In this environment, packaging technologies that offer slightly lower density but higher predictability may outcompete more aggressive approaches that struggle to scale reliably.

From an operational perspective, success will be reflected in how seamlessly optical I/O integrates into datacenter workflows. Maintenance models will adapt to account for deeper optical integration, supported by improved monitoring, diagnostics, and redundancy at the system level. Rather than eliminating failures, successful architectures will localize and tolerate them, ensuring that optical issues do not propagate into widespread system disruption. This shift mirrors broader trends in large-scale computing, where resilience is achieved through architectural design rather than component-level perfection.

Perhaps most tellingly, a successful outcome will be marked by a change in how optical interconnects are discussed. As the technology matures, optics will cease to be framed as an exceptional solution to a crisis in electrical scaling. Instead, optical I/O will be regarded as an expected element of AI system design, subject to the same trade-offs and constraints as other subsystems. The focus will move from asking whether co-packaged optics "work" to determining where, how, and to what extent they should be deployed.

In this sense, success over the next decade will not be defined by the most aggressive integration, the lowest reported energy per bit, or the densest photonic package. It will be defined by architectures that scale predictably, adapt gracefully, and align optical innovation with the realities of AI datacenter operation. Co-packaged optics and 3D photonics will have succeeded not when they push physical limits, but when they become an unremarkable—yet indispensable—part of the system fabric.

## 6. Outlook: questions the field must confront

As co-packaged optics and three-dimensional photonic integration move closer to widespread deployment, the central challenges facing the field are no longer purely technical. Many of the foundational building blocks already exist. What remains unresolved are a set of architectural and systemic questions that will determine whether these technologies mature into robust infrastructure or remain confined to specialized deployments.

A first question concerns thermal coupling. How much thermal interaction between compute and optics is acceptable in practice, and where should boundaries be drawn? While aggressive 3D integration minimizes interconnect length, it also amplifies sensitivity to workload-driven thermal fluctuations. The field must confront whether maximal proximity is always desirable, or whether deliberate separation—enabled by slightly longer but more stable links—offers a better balance at system scale.

A second issue is serviceability and failure semantics. Optical transceivers have historically been among the most replaceable elements in a datacenter. Co-packaged optics challenges this paradigm by embedding optical functionality deep within packages. Should future systems prioritize field-replaceable optical elements, or is the industry prepared to accept higher replacement granularity in exchange for performance gains? If the latter, what architectural mechanisms are required to localize failures and prevent cascading impact?

Standardization raises another set of unresolved questions. At what level should interfaces be standardized to enable a healthy chiplet ecosystem without constraining innovation? Over-standardization risks freezing architectures prematurely, while insufficient standardization perpetuates fragmentation and limits adoption. Striking the right balance will require coordination across device designers, packaging specialists, system architects, and operators—groups that have historically optimized for different objectives.

There is also an open question around metrics of success. Energy per bit has long served as a guiding figure of merit for optical I/O, but its relevance at AI-datacenter scale is increasingly ambiguous. Should future evaluations prioritize thermally stable throughput, yield-adjusted bandwidth, or operational resilience instead? The choice of metrics will shape design priorities and investment decisions, potentially redefining what constitutes progress in the field.

Finally, the field must consider how AI workloads themselves will evolve. Assumptions about communication patterns, synchronization requirements, and accelerator architectures continue to shift. Optical systems optimized for today's training paradigms may prove ill-suited for future inference-dominated or hybrid workloads. Designing optical I/O with sufficient architectural slack to accommodate this uncertainty may be as important as optimizing for current demands.

These questions do not admit simple answers, nor can they be resolved through device innovation alone. They point instead to the need for sustained dialogue across disciplines and across layers of the stack. The next phase of progress in co-packaged optics and 3D photonics will be defined less by isolated breakthroughs than by the collective willingness of the field to confront these tensions directly—and to design systems that embrace, rather than obscure, their inherent trade-offs. As integration deepens, success metrics must evolve from peak link performance toward system-aware measures such as yield-adjusted bandwidth and thermally stable throughput.